\documentclass[%
 reprint,
groupedaddress,
showpacs,preprintnumbers,
amsmath, amssymb, aps,color
]{revtex4-1}

\usepackage{graphicx}% Include figure files
\usepackage{dcolumn}% Align table columns on decimal point
\usepackage{bm}% bold math
\usepackage{color}

\begin{document}

\title{Optical mesh lattices with $\mathcal{PT}$-symmetry}% Force line

\author{Mohammad-Ali Miri$^{1,*}$, Alois Regensburger$^{2,3}$, Ulf Peschel$^{2}$, and Demetrios N. Christodoulides$^{1}$}
\affiliation{$^{1}$CREOL$/$College of Optics, University of Central Florida, Orlando, Florida 32816, USA\\$^{2}$Institute of Optics, Information and Photonics, University of Erlangen-Nuernberg, 91058 Erlangen, Germany\\$^{3}$Max Planck Institute for the Science of Light, 91058 Erlangen, Germany}
\date{\today}
\begin{abstract}
We investigate a new class of optical mesh periodic structures that are discretized in both the transverse and longitudinal directions. These networks are composed of waveguide arrays that are discretely coupled while phase elements are also inserted to discretely control their effective potentials and can be realized both in the temporal and the spatial domain. Their band structure and impulse response is studied in both the passive and parity-time ($\mathcal{PT}$) symmetric regime. The possibility of band merging and the emergence of exceptional points along with the associated optical dynamics are considered in detail both above and below the $\mathcal{PT}$-symmetry breaking point. Finally unidirectional invisibility in $\mathcal{PT}$-synthetic mesh lattices is also examined along with possible superluminal light transport dynamics.
\end{abstract}


\maketitle
\section {INTRODUCTION}

Optical wave propagation in periodic structures has been a theme of considerable attention in the last twenty years or so \cite{b1,a2,a3,a4,a5}. In general such arrangements can exhibit intriguing and potentially useful light dynamics that are otherwise impossible in the bulk. Periodic Bragg gratings and two and three dimensional photonic crystals with a complete band gap are examples of such configurations. Optical waveguide arrays represent yet another class of periodic structures and have been the focus of intense research in the last decade \cite{a6,*a6z1,a7,*a7z1,*a7z2,a8,*a8z1,a9,a10,a11,*a11z1,a12,*a12z1,a13,a14}. As indicated in several studies, optical arrays can be used as versatile platforms to observe a number of processes ranging from Bloch oscillations \cite{a9,a10} to Landau-Zener tunneling \cite{a11}, and from Anderson localization \cite{a12} to discrete solitons \cite{a6,a7} and Rabi oscillations \cite{a13} and dynamic localization \cite{a14}, just to mention a few. Along similar lines, longitudinally periodically modulated optical waveguide arrays have also been studied in several works \cite{a141,*a141z1,a142,*a142z1}.\\

Quite recently, the concept of parity-time ($\mathcal{PT})$ symmetry has been introduced in the field of optics \cite{a15,a16,a17}.  Interestingly, the very idea of $\mathcal{PT}$-symmetry originated within the framework of quantum mechanics through which a large class of complex non-Hermitian Hamiltonians was identified that could in principle exhibit entirely real spectra \cite{a18,*a18z1,*a18z2,a19,*a19z1,*a19z2}. This can happen as long as the associated Hamiltonian and the combined $\mathcal{PT}$ operator share the same set of eigenfunctions. In this case, this is possible provided that the corresponding complex potential satisfies the condition $V^{*}(x)=V(-x)$ \cite{a18}. This directly implies that the real and imaginary parts of the potential must be even and odd functions of position respectively.\\

Lately, optical $\mathcal{PT}$-symmetry has been experimentally observed in two-element coupled systems where non-reciprocal dynamics (disrupting left-right symmetry) and spontaneous $\mathcal{PT}$-symmetry breaking has been demonstrated for the first time \cite{a20,a21}. What made this transition to photonics possible is the isomorphism between the evolution equations in quantum mechanics and optics. Indeed one can show that the complex refractive index, $n(x)=n_{R}(x)+in_{I}(x)$, plays in this case the role of an optical potential. In this representation the real part $n_{R} (x)$ stands for the refractive index profile while the imaginary part $n_{I} (x)$ represents the gain or loss in the system (depending on its sign). Clearly, under $\mathcal{PT}$-symmetric conditions one expects that $n_{R} (x)=n_{R} (-x)$  and $n_{I} (x)=-n_{I} (-x)$. In other words the index distribution must be an even function of position whereas the gain/loss must be anti-symmetric. Thus far, several works have pointed out that $\mathcal{PT}$-symmetry can lead to altogether new optical dynamics which are otherwise impossible in standard passive optical arrangements \cite{a22,a23,a24,a25,a26,a27,a28,a29,a30,a31,a32,a33,a34,a35,a36,a37,*a37z1,a38,a39}. These may include for example the occurrence of abrupt phase transitions along with the appearance of the so-called exceptional points \cite{a22,a23,a24}, power oscillations \cite{a16}, breaking left-right symmetry and the occurrence of secondary emissions \cite{a16}. In addition new classes of optical solitons \cite{a17,a36} and nonlinear $\mathcal{PT}$ optical isolators \cite{a27}, have been suggested along with unidirectional invisibility \cite{a33,a34}, broad area $\mathcal{PT}$ single mode lasers \cite{a28}, and coherent perfect absorbers \cite{a30,a31,a32}. Note that phase transitions similar to those occurring due to $\mathcal{PT}$-symmetry breaking have also been reported in other systems involving gain/loss modulation \cite{a391}.\\

So far however, experimental observations of $\mathcal{PT}$-symmetry in optics have been carried out in basic coupled systems involving only two elements \cite{a20,a21}. What has hindered progress along these lines is not only the delicate balance needed between gain and loss but also the requirement that the real part of the potential should remain symmetric, even in the presence of gain and loss. Fulfilling all these conditions at the same time is by itself a challenging task because of the underlying Kramers-Kronig relations. Therefore of interest would be to develop a new class of optical platforms where refractive effects and gain or loss can be treated separately and thus facilitate the realization of $\mathcal{PT}$-symmetric optics on a large scale. This goal is reached by discretizing not only the transverse, but also the longitudinal or propagation direction.\\

In this work we introduce a new class of $\mathcal{PT}$-symmetric optical lattices. These mesh arrangements are composed of an array of waveguides with each one of them being discretely and periodically coupled to its adjacent neighbors (Fig.~\ref{fig1}(a)). Unlike ordinary waveguide arrays, light propagation in such mesh systems is discretized in two dimensions (transverse and longitudinal). The band structure of this family of mesh lattices is derived analytically and its effects on light dynamics are investigated. Because of the aforementioned 2D-discretization, the resulting band structure is characterized by both a transverse and a longitudinal Bloch momentum.\\

As we will see, this type of lattice can provide a versatile platform for observing a host of $\mathcal{PT}$-symmetric phenomena and processes. Along these lines, phase elements can be readily inserted in the mesh lattice so as to control the real part of the array potential while amplifiers (that are turned on or off) can be included to provide the needed anti-symmetric gain/loss profile. The fundamental building block of such a mesh structure happens to be a basic $\mathcal{PT}$-symmetric coupler arrangement. What makes this structure practically appealing is the physical separation between the coupling and amplification/attenuation stages within this building block. Band merging effects as well as the emergence of exceptional points are investigated in this family of $\mathcal{PT}$ lattices along with superluminal light transport. Finally, unidirectional invisibility in $\mathcal{PT}$-synthetic mesh lattices is also examined and pertinent examples are provided.\\

\section{OPTICAL MESH LATTICES IN THE TIME DOMAIN}

\begin{figure}[t,b,h]
%\begin{center}
\includegraphics[width=3.5in]{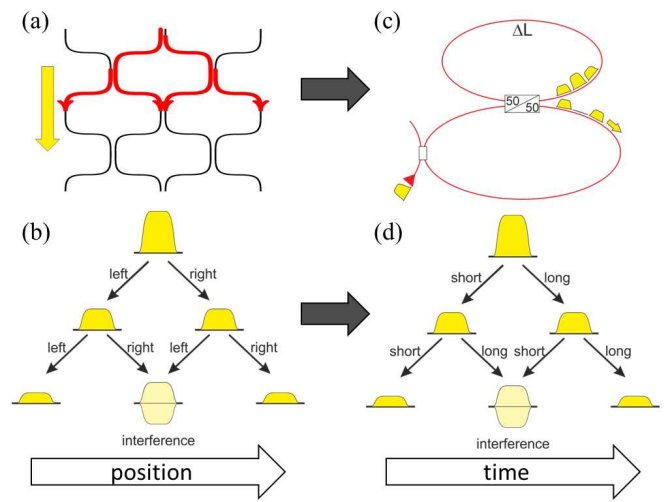}
\caption{(Color online) (a) Discrete mesh lattice in the spatial domain; (b) pulses in the spatial lattice propagate to the left and right on a discrete 1D position grid; (c) equivalent time-multiplexed scheme involving two coupled fiber loops with a length difference $\Delta L$; (d) pulses in the short (long) loop are advanced (delayed) in time, thus making discrete steps on a 1D temporal grid. The propagation dynamics are identical to those expected in (b).}
\label{fig1}
%\end{center}
\end{figure}

Lately, the temporal equivalent of an optical mesh lattice has been experimentally realized using time-multiplexed loop arrangements \cite{a40}. Such configurations have been systematically employed to investigate a number of issues ranging from discrete quantum walks \cite{a41,a42,a43,a44} to Bloch oscillations and fractal patterns \cite{a40,a44bouwmeester}. While spatial realizations of such mesh lattices have also been reported \cite{a44,a45}, time-multiplexed fiber loop schemes have so far demonstrated a high degree of flexibility \cite{a40,a42}. In time-multiplexed schemes, a discrete time axis $n$ corresponds to the transverse discrete axis of a corresponding spatial optical mesh lattice as shown in Figs.~\ref{fig1}(c,d). These time-multiplexed configurations involve two coupled fiber loops-coupled via a central 50/50 directional coupler. These two fiber loops differ in length by $\Delta L$. Here, the equivalent transverse coupling to the left and right sites is enabled by this length difference between two loops.  An independent discretization in time is then obtained by monitoring the round-trip number $m$ in these loops. Hence, the system is essentially discretized in two-dimensions. As we will see, the propagation dynamics of light pulses in such discrete temporal lattices, are exactly identical to those expected in the spatial configurations discussed in section III of this paper.\\

In an experimental set-up, gain and loss can be readily integrated into the fiber loops, e.g. by standard semiconductor or fiber optical amplifiers and amplitude modulators. This in turn may enable the demonstration of large-scale $\mathcal{PT}$-synthetic optical lattices in the temporal domain \cite{a46}. Given that the "topological" arrangements of a temporal and a spatial mesh lattice are totally equivalent, here, without any loss of generality we will consider for simplicity their spatial realization.\\

\section{OPTICAL MESH LATTICE AND ITS BAND STRUCTURE}

\begin{figure}[t,b,h]
%\begin{center}
\includegraphics[width=3.5in]{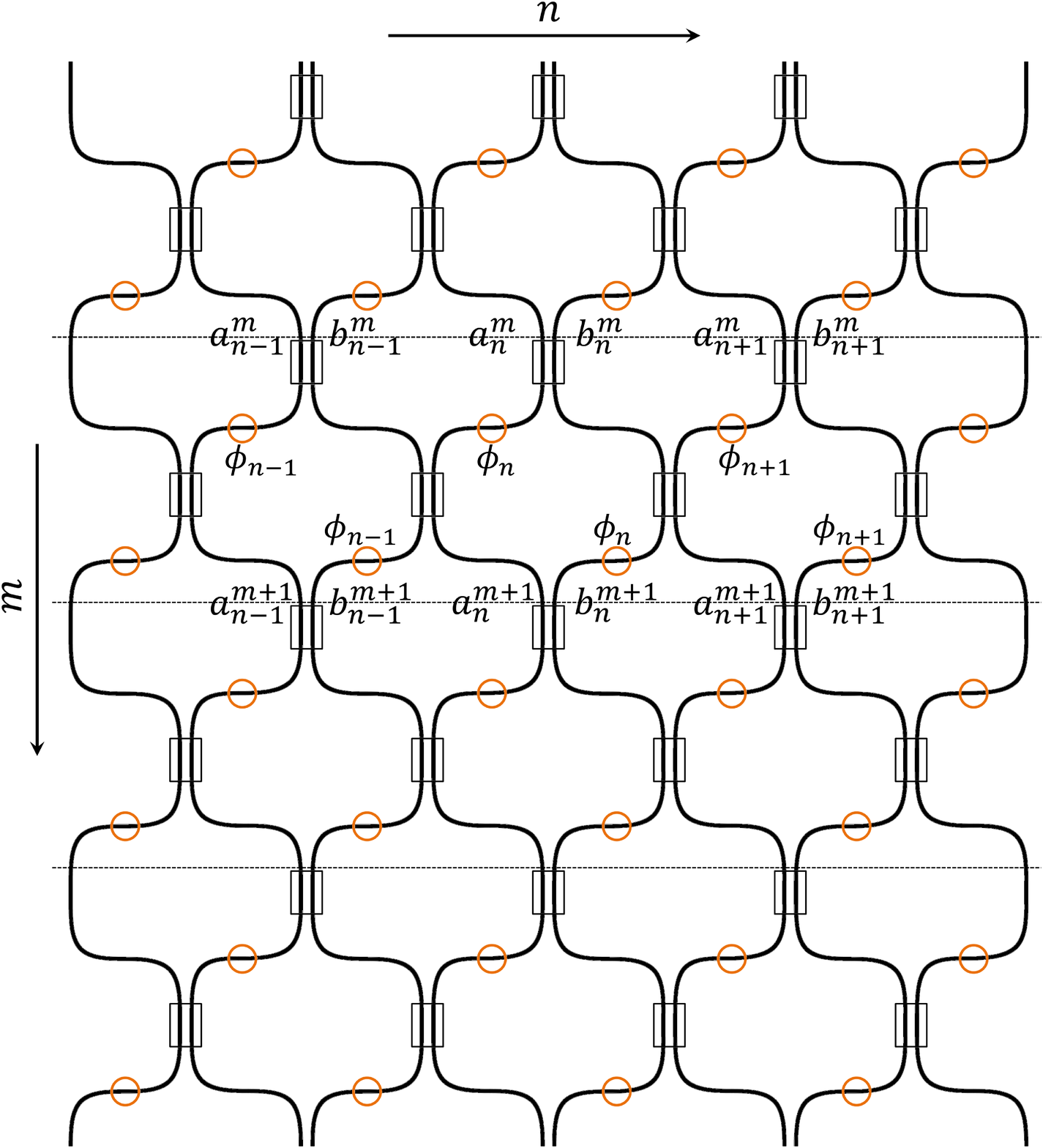}
\caption{(Color online) An optical mesh lattice; The lattice is composed of an array of waveguides which are periodically coupled together in discrete intervals. Circles indicate the position of phase elements and rectangles the coupling regions. The dashed lines show the discrete points where the field intensity is evaluated before coupling occurs.}
\label{fig2}
%\end{center}
\end{figure}

Figure~\ref{fig2} illustrates the spatial realization of such a mesh lattice when only passive phase elements are involved. As previously indicated, this configuration can be synthesized using an array of waveguides that are periodically and discretely coupled to their next neighbors (at the rectangular regions of Fig.~\ref{fig2}). In addition phase elements can also be inserted. Each phase element introduces at every array site $n$ a phase $\phi_{n}$ that happens to be independent of the discrete propagation step $m$. The location of each phase modulator in the lattice is denoted in the figure by a circle.  As we will later demonstrate, these phase modulators effectively play the role of a refractive index profile in continuous arrangements. Figure~\ref{fig1}(a) schematically shows how light flows in such a system when only one of the waveguides is initially excited. After traveling a certain distance in each waveguide, light couples to the adjacent left (right) channel through a coupler, and after propagating this same distance it then couples to the adjacent waveguide to its right (left). Indeed light propagation in this system leads to an interference process that is equivalent to a discrete time quantum walk \cite{a41,a46kempe,a46knight}.\\

As Fig.~\ref{fig2} clearly indicates, this mesh lattice is di-atomic in nature. Using the simple input/output relation of a 50:50 coupler \cite{b47} and by considering the effect of the phase elements, it is straightforward to show that the light evolution equation in this system takes the form:
\small
\begin{subequations}
\label{eq1} % notice location
\begin{eqnarray}
\begin{split}
{a}_{n}^{m+1}=\frac{e^{i\phi_{n}}}{2}[\left({a}_{n}^{m}+i{b}_{n}^{m}\right )\\&+e^{-i\phi_{n}}\left (-a_{n-1}^{m}+ib_{n-1}^{m} \right )], \label{eq1a}
\end{split}
\\
\begin{split}
{b}_{n}^{m+1}=\frac{e^{i\phi_{n}}}{2}[\left({b}_{n}^{m}+i{a}_{n}^{m}\right )\\&+e^{i\phi_{n+1}}\left (-b_{n+1}^{m}+ia_{n+1}^{m} \right )]. \label{eq1b}
\end{split}
\end{eqnarray}
\end{subequations}
\normalsize
In Eqs.~(\ref{eq1}), $a_{n}^{m}$ and $b_{n}^{m}$ represent the field amplitudes  at adjacent waveguide sites $n$ (in the $n$'th column) at a discrete propagation step or distance  $m$ ( $m$'th row). It should be noted that in deriving these equations the phase accumulated due to propagation in any waveguide section is ignored. Indeed a waveguide section of length $l$ between two subsequent couplers leads to a phase accumulation of $\beta l$, where $\beta$ is the propagation constant of the guide. Yet, one can readily show that even in the presence of these additional phase terms Eqs.~(\ref{eq1}) remain the same once a simple gauge transformation is used; $(a_{n}^{m},b_{n}^{m} )\rightarrow(a_n^m,b_n^m ) e^{i2m\beta l}$.\\

To establish the necessary periodicity, we assume that the phase elements provide a phase potential that alternates between two different values in $n$:
\begin{equation}
\label{eq2}
\phi_{n}=
\begin{cases}
+\phi_{0},~~~n:~ even\\
-\phi_{0},~~~n:~~ odd
\end{cases}
\end{equation}
This kind of phase potential has a translational symmetry $\phi_{n+2}=\phi_n$ which leads to a transverse periodicity in this "four-atom" lattice with a fundamental period of $N=2$ where each cell is diatomic. In addition the lattice is now periodic in both $n$ and $m$.\\

First we study the band structure of this mesh system. Once the band characteristics and corresponding Bloch modes are known, the dynamic properties of the system can then be extrapolated. To find the dispersion relation of this lattice we consider discrete "plane wave solutions" of the form $e^{iQn} e^{i\theta m}$ where $Q$ represents a Bloch momentum in the transverse direction and $\theta$ plays the role of a propagation constant. To obtain the corresponding band structure we assume solutions of the form:
\begin{equation}
\label{eq3}
\binom {a_{n}^{m}}{b_{n}^{m}}=\binom {A_{n}}{B_{n}}e^{iQn} e^{i\theta m}
\end{equation}
where $A_{n}$ and $B_{n}$ are periodic Bloch functions with the period of $N=2$, i.e. $A_{n+2}=A_{n}$ and $B_{n+2}=B_{n}$. In general, for  $n=2j$, we use $A_n,B_n=A_0,B_0$ while for $n=2j+1$ we employ $A_n,B_n=A_1,B_1$. This comes from the fact that a unit cell of this periodic structure includes two discrete positions $n$.\\

\begin{figure}[t,b,h]
%\begin{center}
\includegraphics[width=3.5in]{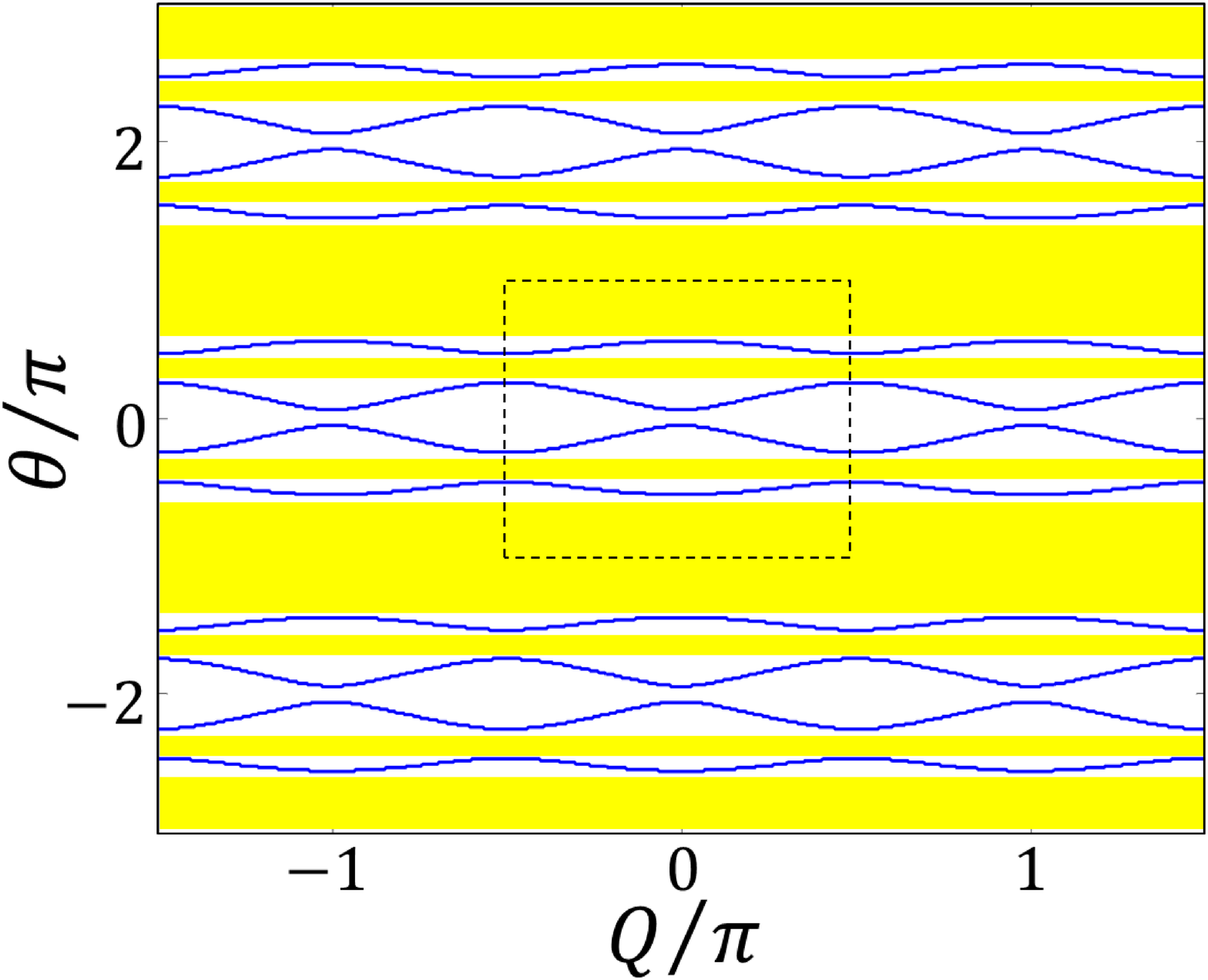}
\caption{Band structure of the optical mesh lattice in the presence of periodic step-like potential created from phases, alternating between $-\phi_{0}$ and $+\phi_{0}$ where $\phi_{0}=0.2\pi$. The shaded  area shows the band gap regions and the dotted boundary depicts the primary Brillouin zone of this lattice.}
\label{fig3}
%\end{center}
\end{figure}

By inserting Eqs.~(\ref{eq3}) in (\ref{eq1}), and by adopting the phase potential of Eq.~(\ref{eq2}), we obtain the following dispersion relation after expanding the corresponding $4\times 4$ determinant of a unit cell:
\begin{equation}
\label{eq4}
cos(2Q)=8{cos}^{2}(\theta)-8cos(\phi_{0})cos(\theta)+4{cos}^{2}(\phi_{0})-3
\end{equation}
As expected from the double periodicity of this system in both $n$ and $m$ the band structure is also periodic in both $Q$ and $\theta$ having fundamental periods of $\pi$ and $2\pi$ respectively. This represents a major departure from optical waveguide arrays where the propagation dimension is a continuous variable. Under the assumption of Eq.~(\ref{eq2}), this mesh arrangement exhibits four primary bands which are periodic with respect to the two Bloch momenta. Figure~\ref{fig3} depicts the band structure of this mesh lattice when $\phi_{0}=0.2\pi$.

\begin{figure}[t,b,h]
%\begin{center}
\includegraphics[width=3.5in]{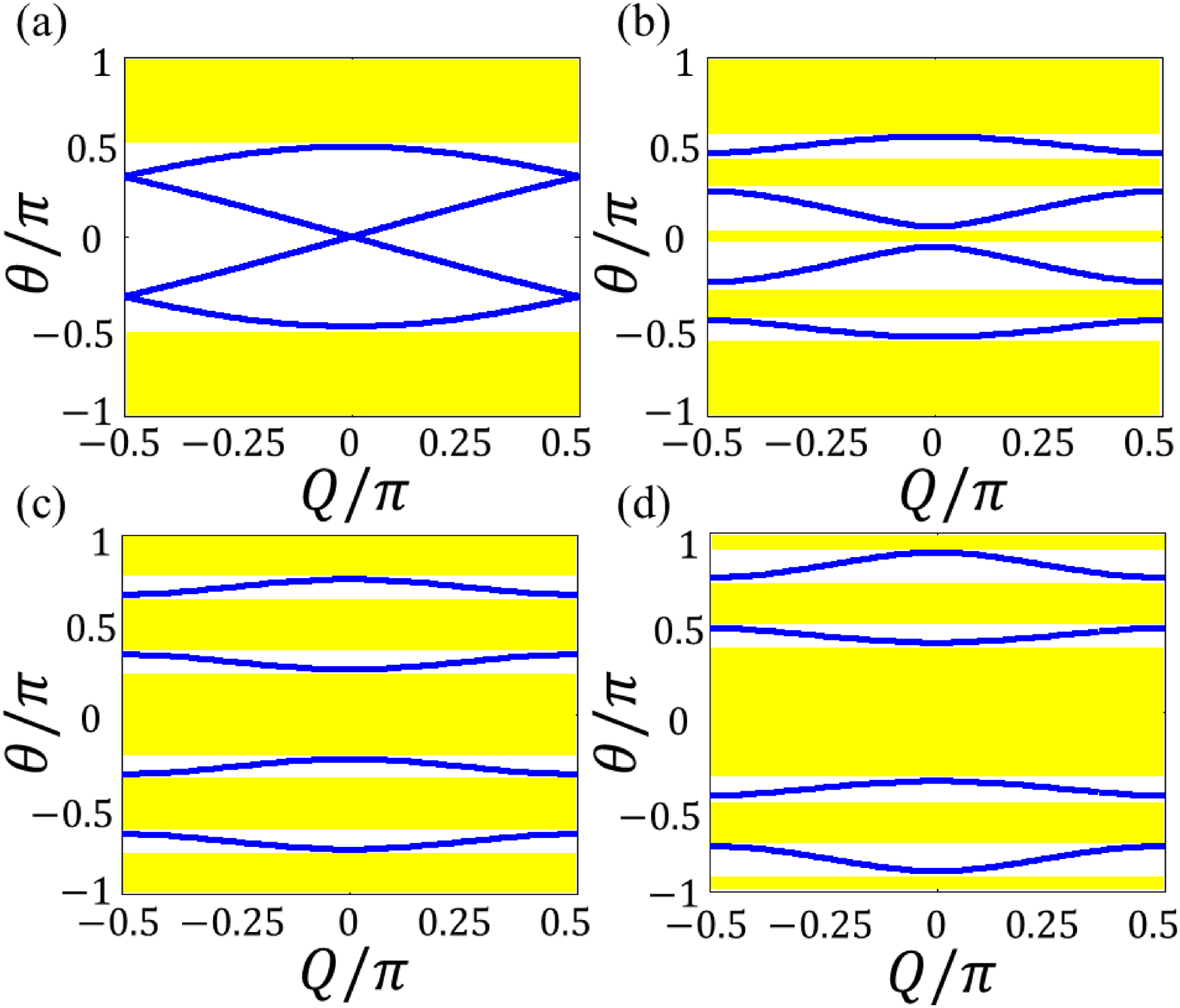}
\caption{(Color online) Band structure of an optical mesh lattice for several cases; (a) Lattice without any phase potential $\phi_{0}=0$ (empty lattice), (b) Lattice with a symmetric phase step-like potential varying between $-\phi_{0}$ and $+\phi_{0}$ when $\phi_{0}=0.2\pi$, (c) same as in (b) but with $\phi_{0}=0.5\pi$,  (d) $\phi_{0}=0.7\pi$. For case (a) the reduced Brillouin zone is depicted while for the rest the first Brillouin zone is shown in its entirety.}
\label{fig4}
%\end{center}
\end{figure}

Equation~\ref{eq4} is valid in general for any arbitrary choice of $\phi_{0}$. However it should be noticed that in the special case where $\phi_{0}=0$ (empty lattice) this relation becomes degenerate. Indeed for the empty lattice the periodicity of this diatomic lattice is $N=1$ and hence its Brillouin zone involves two bands and lies in the domain of $-\pi<Q<\pi$ and $-\pi<\theta<\pi$. The folded version of this Brillouin zone (corresponding to the empty lattice) is shown in Fig.~\ref{fig4}(a) where the two bands are degenerately folded into four. Figures~\ref{fig4}(b,c,d) depict the band structure of this mesh lattice for three nonzero values of $\phi_{0}$ within the Brillouin zone as a function of the Bloch momenta, i.e., $-\pi/2<Q<\pi/2$ and $-\pi<\theta<\pi$. Again the shaded areas show the associated band gaps. According to Fig.~\ref{fig4}, a nonzero $\phi_0$ lifts the degeneracy and leads indeed to four bands.

According to Eq.~(\ref{eq4}) and as one can see from the figures the band structure has a reflection symmetry around $Q=0$ and $\theta=0$. For any finite $\phi_{0}$ there are four bands in the Brillouin zone, all having a zero slope at the center ($Q=0$) and at the edges ($Q=±\pi/2$). For the empty lattice on the other hand, in reality there are two bands and the slope is zero at the center ($Q=0$) of the top band while it is non-zero at the two edges ($Q=±\pi/2$) and at $Q=\theta=0$ where the bands collide and there is no band gap between them. The addition of the phase potential $±\phi_{0}$ to the empty lattice breaks this degeneracy and creates band gaps at these points. This breaking of the degeneracy becomes clear by comparing Figs.~\ref{fig4}(a) and (b). Equation~\ref{eq4} can also be written in a more explicit form as a function of $Q$:
\small
\begin{equation}
\label{eq5}
\theta=\pm {cos}^{-1}\left[\frac{1}{2}\left(cos(\phi_{0})\pm\sqrt{{cos}^{2}(Q)+{sin}^{2}(\phi_{0})}\right)\right]
\end{equation}
\normalsize
where in this relation any combination of the two plus/minus signs corresponds to each of the four bands.\\

\begin{figure}[t,b,h]
%\begin{center}
\includegraphics[width=3.5in]{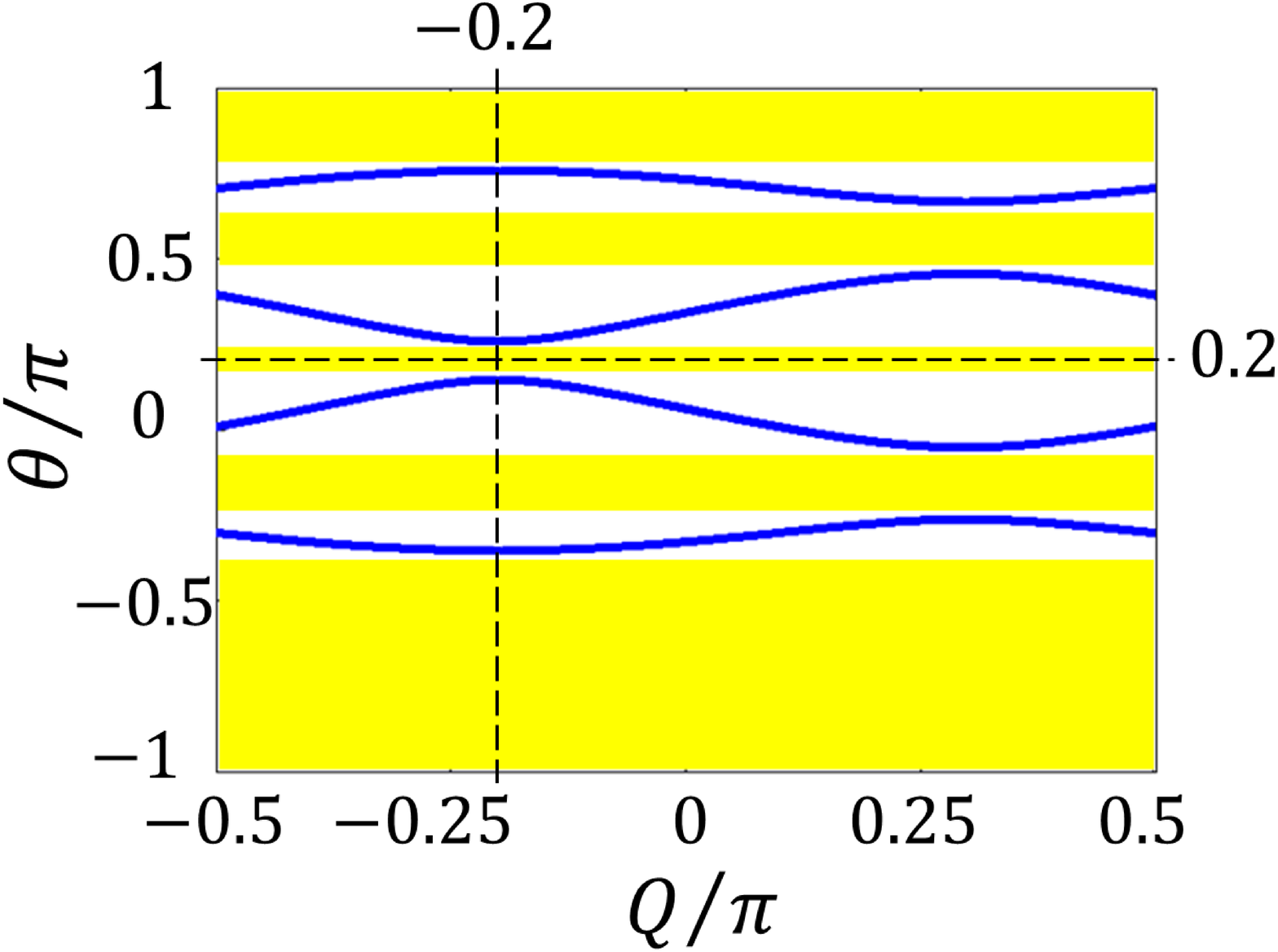}
\caption{(Color online) Band structure of an optical mesh lattice with a non-symmetric step-like phase potential alternating between $0$ and $2\phi_{0}$ while $\phi_{0}=0.2\pi$. Compared to the case of a symmetric phase potential (Fig.~\ref{fig4}(b)) the band structure is shifted from the center.}
\label{fig5}
%\end{center}
\end{figure}

Before ending this discussion, it is worth noting that this phase potential does not need to be symmetrized in a $±\phi_{0}$ fashion as done before in this section. In fact any periodic potential that is alternating in $n$ between two different phase values will break the degeneracy of an empty lattice, thus creating four bands in the first Brillouin zone. For example let us consider a phase potential that varies between $0$ and $2\phi_{0}$ in $n$:
\begin{equation}
\label{eq6}
\phi_{n}=
\begin{cases}
2\phi_{0},~~~n:~ even\\
0,~~~~~~n:~~ odd
\end{cases}
\end{equation}
Note that this latter phase potential has the same strength as the one used before. In this latter case, by using the same ansatz of Eq.~(\ref{eq3}) we directly obtain the dispersion relation corresponding to the new potential of Eq.~(\ref{eq6}).
\small
\begin{equation}
\label{eq7}
\begin{split}
cos\left(2(Q+\phi_{0})\right)=8{cos}^{2}(\theta-\phi_{0})-8cos(\phi_{0})cos(\theta-\phi_{0})\\+4{cos}^{2}(\phi_{0})-3
\end{split}
\end{equation}
\normalsize
A close examination of Eq.~(\ref{eq7}) reveals that this latter dispersion curve is identical to that of Eq.~(\ref{eq4}), apart from a phase shift in both $\theta$ and $Q$. More specifically $Q$ has shifted by an amount of  $-\phi_{0}$ while $\theta$ by $\phi_{0}$. Figure~\ref{fig5} shows a plot of this dispersion relation for $\phi_{0}=0.2\pi$. The shift of origin compared to Fig.~\ref{fig4}(b) is evident in this figure.

In the rest of this work we consider for simplicity symmetric phase potentials for which the band structure is symmetric around $Q=\theta=0$.\\

\section{OPTICAL DYNAMICS IN MESH LATTICES}

In this section we investigate optical dynamics in passive mesh lattices. The impulse response of the system is of particular importance since is known to excite the entire band structure. For this reason only one of the waveguide elements is excited at $m=0$. In what follows, the impulse response will be studied by using ${a}_{0}^{0}=1$ with all the other elements in the array initially set to zero.

\begin{figure}[t,b,h]
%\begin{center}
\includegraphics[width=3.5in]{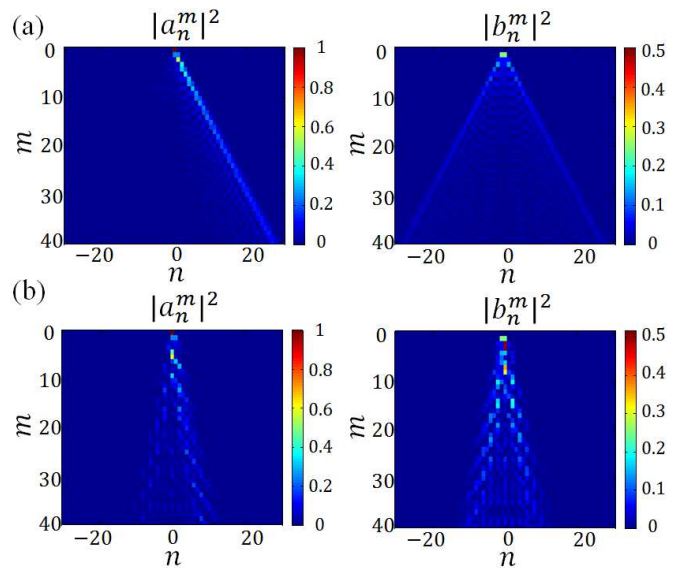}
\caption{(Color online) Impulse response of a mesh lattice where the intensity profile of ${a}_{n}^{m}$ and ${b}_{n}^{m}$ is plotted; (a) $\phi_{0}=0$ (empty lattice), (b) $\phi_0=0.4\pi$. In both cases ${a}_{0}^{0}=1$ and all other elements are initially set to zero.}
\label{fig6}
%\end{center}
\end{figure}

Figure~\ref{fig6}(a) shows the impulse response of this array lattice when $\phi_{0}=0$. According to this figure light transport in this system exhibits upon spreading a highest slope of $\Omega_{max}=\pm 1/\sqrt{2}$ with respect to the longitudinal axis. As we will see this result will be formally justified by considering the group velocity in this arrangement. The impulse response of the mesh lattice in the presence of a periodic phase potential with $\phi_{0}=0.4\pi$ is also plotted in Fig.~\ref{fig6}(b). In this last case, it becomes clearly apparent that the maximum speed of the excitation spreading becomes slower when $\phi_{0}$ increases. As in waveguide arrays \cite{a6}, the impulse response can be viewed as a "ballistic" transport across the array.\\

The band structure can also provide useful information concerning the evolution of more complicated initial excitations like localized wavepackets. More specifically, we consider initial distributions of ${a}_{n}^{0}$ and ${b}_{n}^{0}$ of the form $f_{n} e^{iQ_{0} n}$ where $f_{n}$ is a slowly varying envelope function (with a narrow spatial spectrum) and $e^{iQ_{0} n}$ is a rapidly varying phase term signifying the central Bloch momentum $Q_{0}$ of this wavepacket. Therefore the propagation process of this discrete beam excitation can be effectively treated through a Fourier superposition of the Floquet-Bloch modes $e^{iQn} e^{i\theta m}$ assumed before to analyze this system. In this regard, both the group velocity and the dispersion broadening of this wavepacket can be obtained  by expanding the propagation constant $\theta$ in a Taylor series around $Q_{0}$, that is:
\begin{equation}
\label{eq8}
\theta=\theta_0+\frac{\mathrm{d}\theta}{\mathrm{d}Q}~_{{|}_{Q_{0}}}(Q-Q_{0})+\frac{{\mathrm{d}}^{2}\theta}{\mathrm{d}Q^{2}}~_{{|}_{Q_{0}}}(Q-Q_{0})
\end{equation}
As in continuous lattices, the tangent of the beam angle (or "group velocity") is associated with the term:
\begin{equation}
\label{eq9}
\Omega=\frac{\mathrm{d}\theta}{\mathrm{d}Q}~_{{|}_{Q_{0}}}
\end{equation}
Using the dispersion Eq.~(\ref{eq4}), this group speed can then be written as:
\begin{equation}
\label{eq10}
\frac{\mathrm{d}\theta}{\mathrm{d}Q}=\frac{1}{4}\frac{sin(2Q)}{\left [sin(2\theta)-cos(\phi_{0})sin(\theta)\right ]}
\end{equation}
where in this relation $\theta$ could be replaced from the dispersion relation of Eq.~(\ref{eq5}) to obtain the right hand side as a function of $Q$ and the band under consideration. Using similar arguments, the discrete diffraction factor can be obtained from:
\begin{equation}
\label{eq11}
D=\frac{{\mathrm{d}}^{2}\theta}{\mathrm{d}{Q}^{2}}~_{{|}_{Q_{0}}}
\end{equation}

\begin{figure}[t,b,h]
%\begin{center}
\includegraphics[width=3.5in]{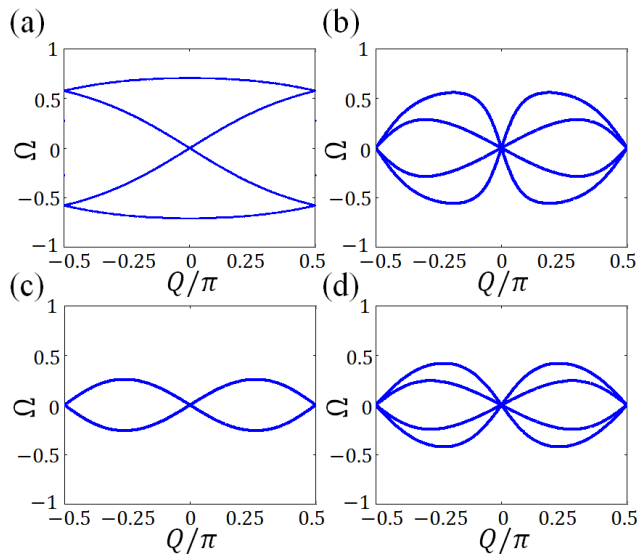}
\caption{(Color online) Beam tangent angle ($\Omega$) for several cases; (a) empty lattice (note that in this case the curve is folded to the reduced Brillouin zone), (b) for a lattice in the presence of periodic phase potential with $\phi_{0}=0.2\pi$, (c) $\phi_{0}=0.5\pi$, (d) $\phi_{0}=0.7\pi$.}
\label{fig7}
%\end{center}
\end{figure}

Figure~\ref{fig7} depicts the beam angle $\Omega$ for several lattices with different amplitudes of the phase potential, $\phi_{0}$. According to this figure, in an empty lattice ($\phi_{0}=0$) this beam angle is zero at the center ($Q=0$) and it is maximum at $Q=\theta=0$ in the folded Brillouin zone scheme where to first order the dispersion relation dictates that $Q=\pm \sqrt{2}\theta$. Hence, as previously indicated, the maximum slope expected in this configuration is $\Omega_{max}=\pm 1/\sqrt{2}$. On the other hand for a lattice having a periodic phase potential, each band exhibits a zero group velocity at the center and at the edges ($Q=\pm \pi/2$) of the zone while the maximum happens somewhere in between. For the special case of $\phi_{0}=\pi/2$  the bands are translated in $\theta$ and hence in groups of two have identical group velocity curves, and as shown in Fig.~\ref{fig7}(c) they lie on top of each other.

\begin{figure}[t,b,h]
%\begin{center}
\includegraphics[width=3.5in]{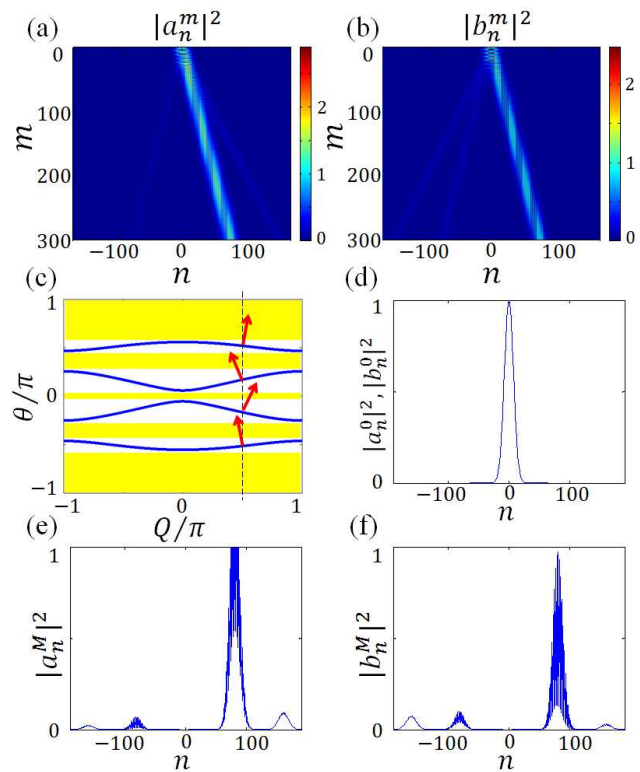}
\caption{(Color online) Gaussian wavepacket propagating in a mesh lattice. The beam has a width of $2\Delta=30$ and an initial phase tilt of $Q_0=0.25\pi$. The lattice has a phase potential of $\phi_0=0.2\pi$ (a) intensity $\left |{a}_{n}^{m}\right |^2$, (b) intensity of $\left |{b}_{n}^{m}\right |^2$, (c) band structure of the lattice with the dashed line crossing the band at four points at $Q_0=0.25\pi$ and the arrows show the propagation direction of the four resulting beams, (d) intensity profile of the initial Gaussian beam, (e) $\left |{a}_{n}^{M}\right |^2$ intensity profile of ${a}_{n}^{m}$ at the last discrete longitudinal step (here $M=300$), (e) $\left |{b}_{n}^{M}\right |^2$.}
\label{fig8}
%\end{center}
\end{figure}

To demonstrate some of these transport effects, let us consider for example the evolution of a Gaussian wavepacket having the following initial profile:
\begin{equation}
\label{eq12}
{a}_{n}^{0}=e^{-{(n/\Delta)}^{2}} e^{iQ_{0}n}
\end{equation}
where $2\Delta$ represents the Gaussian beamwidth and $Q_0$ designates the initial tilt in its phase front or central Bloch momentum. In this case the same input profile is assumed for $b_{n}^{0}$ in order to symmetrize the dynamics. Figure~\ref{fig8} shows the propagation dynamics of this Gaussian beam in this mesh lattice. Here the lattice involves a periodic phase potential with $Q_0=0.2\pi$. The Gaussian beam width $2\Delta$ is large enough to avoid the diffraction effects and in addition its tilt is $Q_0=0.25\pi$. According to this figure four independent beams (of the same width) result from this initial excitation, each emanating from a corresponding band, and propagating in different directions. To elucidate these results, the band structure is also plotted in this same Fig.~\ref{fig8}(c) where the arrows perpendicular to the bands indicate the propagation direction of each of these four beams.

\begin{figure}[t,b,h]
%\begin{center}
\includegraphics[width=3.5in]{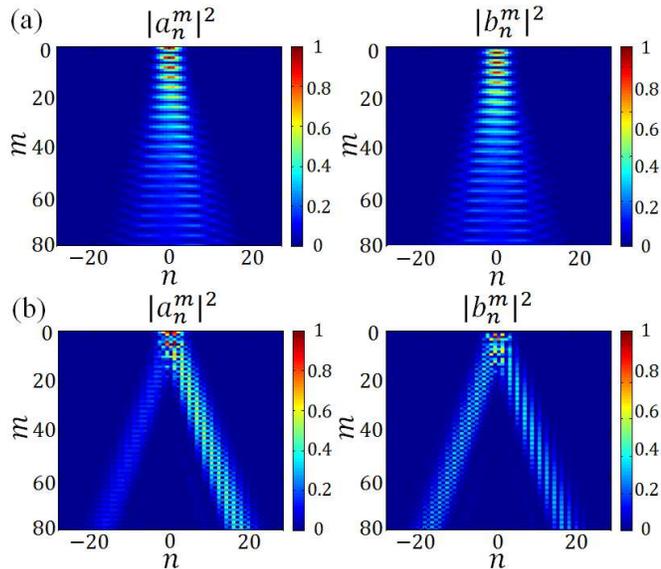}
\caption{(Color online) Diffraction properties of a Gaussian beam in a mesh lattice with $\phi_0=0.5\pi$. The Gaussian beam has a width of $2\Delta=8$ while the initial phase tilt is: (a) $Q_0=0$, (b) $Q_0=0.25\pi$.}
\label{fig9}
%\end{center}
\end{figure}

Finally in order to investigate diffraction effects in passive mesh systems, we consider the propagation properties of a relatively narrow Gaussian wavepacket. Figure~\ref{fig9} depicts the propagation dynamics of a Gaussian beam with a width of $2\Delta=8$ in a lattice with $\phi_0=0.5\pi$. The figures compare the beam propagation for two different values of $Q_0$, $0$ and $0.25\pi$. According to this figure when $Q_0=0$, the beam has a very low transverse velocity and experiences a considerable degree of diffraction. As shown in the other panels, when the beam is launched at the dispersion free point of the band ($Q_0=0.25\pi$), where $D=0$ and the transverse group velocity is maximum, the diffraction effects are negligible. 

According to Fig.~\ref{fig4} this selection of $\phi_0$ leads to four bands.  Figure~\ref{fig9}(a) depicts Gaussian beam spreading at $Q_0=0$ and at the same time interference effects resulting from the excitation of multiple bands. On the other hand for $Q_0=0.25\pi$ two Gaussian beams symmetrically emerge with two different propagation speeds. Yet, the interference pattern in each of the two branches demonstrates that all four bands are actually in play in these dynamics. Notice however that at this point little beam spreading occur since for these parameters $D=0$.\\

\section{$\mathcal{PT}$-SYMMETRIC BUILDING BLOCK}

\begin{figure}[t,b,h]
%\begin{center}
\includegraphics[width=3.5in]{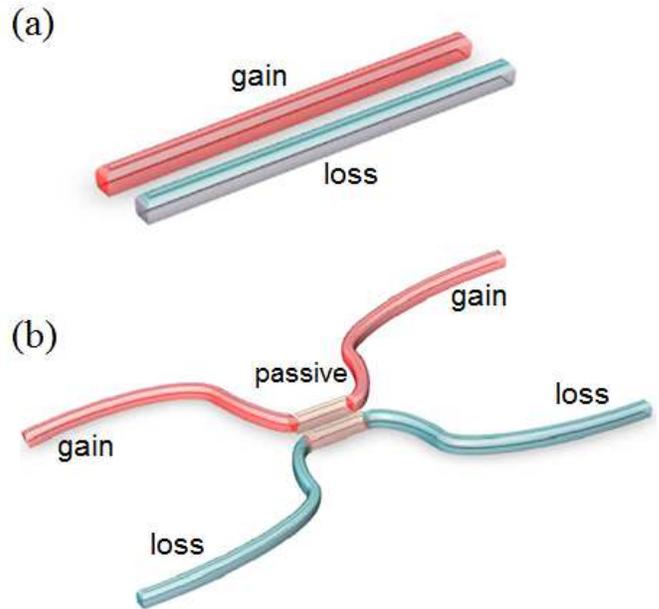}
\caption{(Color online) A distributed $\mathcal{PT}$-symmetric coupler and a $\mathcal{PT}$-synthetic coupler; (a) The $\mathcal{PT}$-coupler is composed of two similar dielectric waveguides coupled to each other, with one experiencing gain (red) while the other an equal amount of loss (blue), (b) A $\mathcal{PT}$-synthetic coupler is composed of a passive coupler while the gain and loss waveguides are separately used in the arms.}
\label{fig10}
%\end{center}
\end{figure}

Before exploring a large-scale $\mathcal{PT}$-symmetric mesh lattice, it is worth analyzing the elemental building block involved in such a network. Figure~\ref{fig10}(a) shows a $\mathcal{PT}$-symmetric coupler where the gain and loss is uniformly distributed along the two arms, a structure similar to that considered in previous experimental studies \cite{a20, a21}. Figure~\ref{fig10}(b) on the other hand depicts a passive coupler where the gain and loss mechanisms are separately inserted in the two arms only. Here we show that this new type of $\mathcal{PT}$-symmetric coupler displays exactly the same behavior and characteristics of a standard $\mathcal{PT}$-coupler arrangement considered before.

In Figure~\ref{fig10}(b) we assume a 50:50 passive directional coupler connected to two arms, one providing amplification (red) while the other an equal amount of loss (blue). We assume that each arm delivers an amplification or attenuation of $e^{\pm \gamma/2}$ right before and after the coupler. Hence the modal amplitudes $a'$ and $b'$ at the output of this system, are related to those at the input ports, $a$ and $b$, in the following way:
\small
\begin{equation}
\label{eq13}
\binom {a'}{b'}=
\begin{pmatrix} e^{+\gamma/2} & 0 \\ 0 & e^{-\gamma/2} \end{pmatrix}
\frac{1}{\sqrt{2}}
\begin{pmatrix} 1 & i \\ i & 1 \end{pmatrix}
\begin{pmatrix} e^{+\gamma/2} & 0 \\ 0 & e^{-\gamma/2} \end{pmatrix}
\binom {a}{b} 
\end{equation}
\normalsize
in which case
\begin{equation}
\label{eq14}
\binom {a'}{b'}=
\frac{1}{\sqrt{2}}
\begin{pmatrix} e^{+\gamma} & i \\ i & e^{-\gamma} \end{pmatrix}
\binom {a}{b}
\end{equation}
where $a$ and $b$ represent optical amplitudes in the gain and loss channels respectively. The two supermodes and their respective eigenvalues of this system can be readily found. Depending on the amount of gain/loss in the system two regimes can be distinguished; if $\gamma<{cosh}^{-1}(\sqrt{2})$ this $\mathcal{PT}$ system is operating below the $\mathcal{PT}$-symmetry breaking threshold and its supermodes are given by:
\begin{equation}
\label{eq15}
\binom {a_0}{b_0}=\binom {1}{\pm e^{\pm i\eta}}e^{\pm i\omega} 
\end{equation}
where $cos(\omega)=\frac{1}{\sqrt{2}} cosh(\gamma)$ and $sin(\omega)=\frac{1}{\sqrt{2}} cos(\eta)$. Thus for $\gamma<{cosh}^{-1}(\sqrt{2})$ the two modes repeat themselves after passing through this discrete system except from a trivial phase shift of $\pm \omega$. On the other hand if $\gamma>{cosh}^{-1}(\sqrt{2})$ the system operates above the $\mathcal{PT}$-symmetry breaking threshold and:
\begin{equation}
\label{eq16}
\binom {a_0}{b_0}=\binom {1}{i e^{\mp \eta}}e^{\pm \omega} 
\end{equation}
where $cosh(\omega)=\frac{1}{\sqrt{2}} cosh(\gamma)$ and $sinh(\omega)=\frac{1}{\sqrt{2}} sinh(\eta)$. Interestingly this same behavior is displayed by a standard $\mathcal{PT}$-symmetric coupler where the gain and loss is continuously distributed. Finally at exactly the $\mathcal{PT}$-symmetry breaking threshold $\gamma={cosh}^{-1}(\sqrt{2})$ the two supermodes collapse to one and thus:
\begin{equation}
\label{eq17}
\binom {a_0}{b_0}=\binom {1}{i} 
\end{equation}
which clearly shows the existence of a phase difference of $\pi/2$ between the two waveguides.\\

It is worth noting that this arrangement has certain advantages over a standard distributed $\mathcal{PT}$-symmetric coupler. First of all it is experimentally easier to achieve the delicate balance required for $\mathcal{PT}$ symmetry. In addition the coupling and amplification/attenuation process take place in two separate steps so there are no physical restrictions imposed by the Kramers-Kronig relations. As previously mentioned, these effects have so far hindered progress in implementing large-scale $\mathcal{PT}$-symmetric networks, since they limit the possibility of achieving the required values for gain/loss and refractive index, all at the same time.\\

\section{$\mathcal{PT}$-SYNTHETIC MESH NETWORKS}

Figure~\ref{fig11}(a) shows a $\mathcal{PT}$-symmetric mesh lattice made of $\mathcal{PT}$-synthetic couplers, identical to that of Fig.~\ref{fig10}(b). In addition phase elements are inserted in this same lattice (shown by circles in Fig.\ref{fig11}(a)) in order to provide the needed real part in the potential function. In Fig.~\ref{fig11}(b) the distributions of phase modulation and that of gain/loss are plotted as a function of the discrete position $n$ - clearly satisfying the requirement for $\mathcal{PT}$-symmetry, i.e. an even distribution for the phase and an odd distribution for the gain/loss profile in $n$. In fact a comparison with continuous systems suggests that the phase and gain/loss in discrete elements play the role of the real and imaginary parts in the refractive index respectively. By considering an amplification/attenuation factor of $e^{\pm \gamma/2}$ in each waveguide section between two subsequent couplers, then one can show that light propagation in this $\mathcal{PT}$-synthetic mesh network is governed by the following discrete evolution equations:

\begin{figure}[t,b,h]
%\begin{center}
\includegraphics[width=3.5in]{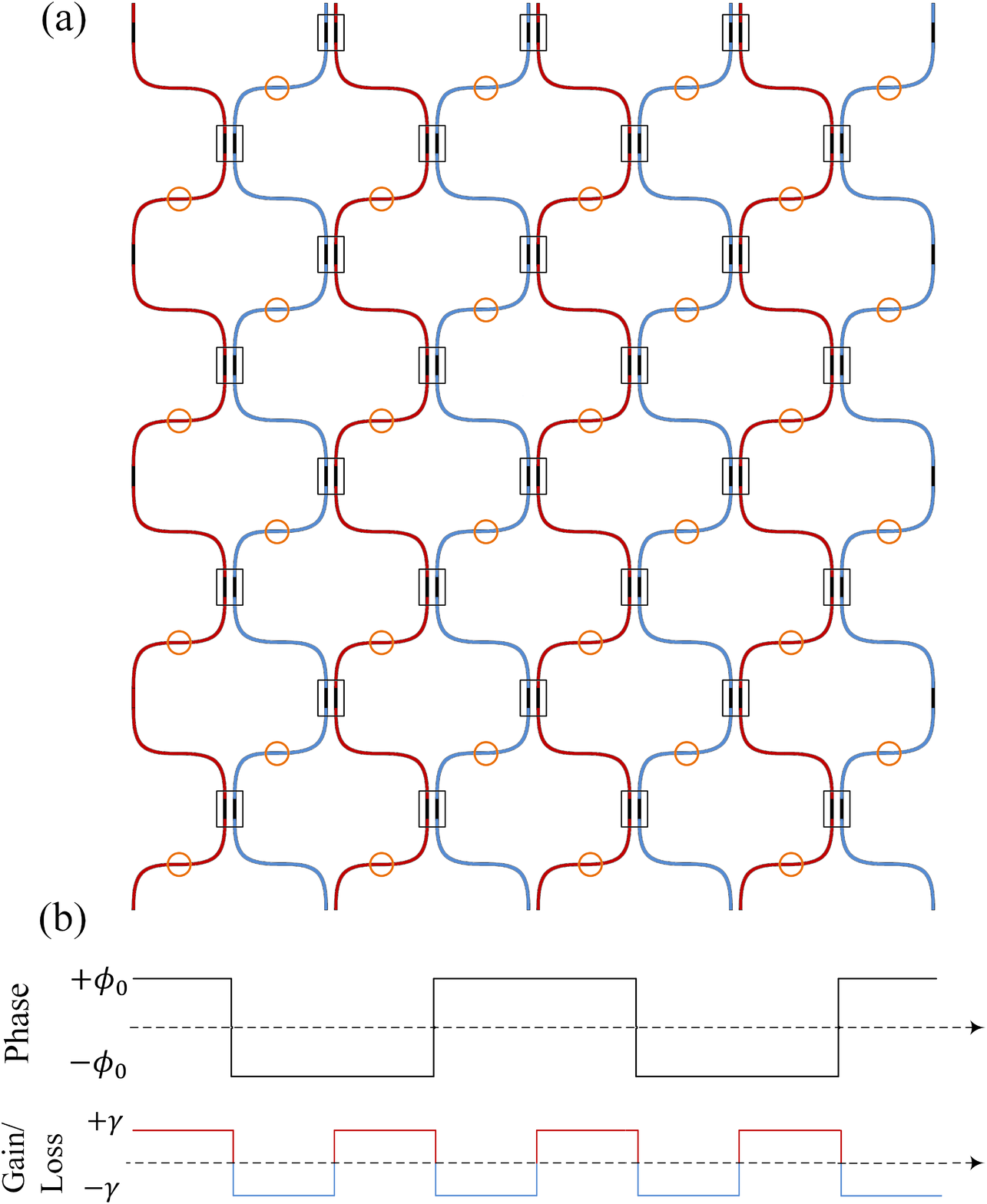}
\caption{(Color online) (a) A $\mathcal{PT}$-synthetic mesh lattice, (b) transverse distribution of the phase potential (symmetric) and gain/loss (antisymmetric).}
\label{fig11}
%\end{center}
\end{figure}

\small
\begin{subequations}
\label{eq18} % notice location
\begin{eqnarray}
\begin{split}
{a}_{n}^{m+1}=\frac{e^{i\phi_{n}}}{2}[e^{-\gamma}\left({a}_{n}^{m}+i{b}_{n}^{m}\right )\\&+e^{-i\phi_{n}}\left (-a_{n-1}^{m}+ib_{n-1}^{m} \right )], \label{eq1a}
\end{split}
\\
\begin{split}
{b}_{n}^{m+1}=\frac{e^{i\phi_{n}}}{2}[e^{+\gamma}\left({b}_{n}^{m}+i{a}_{n}^{m}\right )\\&+e^{i\phi_{n+1}}\left (-b_{n+1}^{m}+ia_{n+1}^{m} \right )]. \label{eq1b}
\end{split}
\end{eqnarray}
\end{subequations}
\normalsize

\begin{figure}[t,b,h]
%\begin{center}
\includegraphics[width=3.5in]{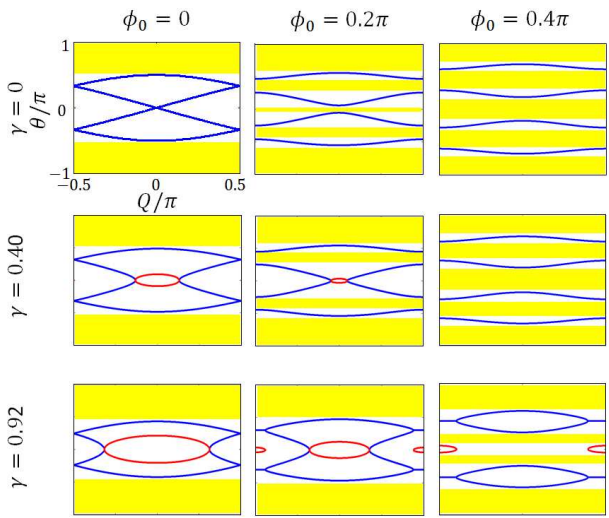}
\caption{(Color online) Band structure of $\mathcal{PT}$-synthetic mesh lattice for several values of $\phi_0$ and $\gamma$. In these plots the real part of propagation constant, $\theta$ is indicated in blue, while the imaginary part in red.}
\label{fig12}
%\end{center}
\end{figure}

To understand the behavior of this system, the band structure should be first determined. By adopting the same ansatz of Eq.\ref{eq3}, one can derive the following dispersion relation for this $\mathcal{PT}$ lattice:
\begin{equation}
\label{eq19}
\begin{split}
cos(2Q)=8{cos}^{2}(\theta)-8cosh(\gamma)cos(\phi_{0})cos(\theta)\\+4{cos}^{2}(\phi_{0})-4+cosh(2\gamma)
\end{split}
\end{equation} 
Figure~\ref{fig12} shows the band structure of this system for several different values of the phase potential amplitude $\phi_0$ and gain/loss coefficients $\gamma$. In each case the real parts of the propagation constant ($\theta$) is plotted in blue while the imaginary parts are shown in red.

As it is illustrated in this figure, the presence of a symmetric phase potential in this system tends to pull apart the bands thus creating a band gap, while the antisymmetric gain/loss tends instead to close the gap. The system is said to be operating below the $\mathcal{PT}$-symmetry breaking threshold as long as the eigenvalues associated with all bands are real. However at a critical amount of gain/loss the bands merge at the so called \it exceptional points\rm, and for even higher gain/loss values, sections with conjugate imaginary eigenvalues appear in the bands.\\

In what follows, we consider the case where $\phi_0$ is fixed and discuss how the band structure will change by gradually increasing the gain/loss coefficient $\gamma$. Analysis shows, that for a given value of $\phi_0$, the first band merging occurs at two different positions; if $0<\phi_0<\pi/4$, the bands merge at $Q=\theta=0$ and the second band gap remains open till reaching a critical value of gain/loss coefficient $\gamma$. For even higher gain/loss values the system finds itself in the broken phase regime. For $\pi/4<\phi_0<\pi/2$ on the other hand all bands are open till a critical point. Exactly at this threshold, the band gap at the edges of the Brillouin zone at $Q=\pm \pi/2$ closes while the first band gap remains open till reaching another critical point where it eventually evaporates. Based on these observations analytical results for the symmetry breaking point can be obtained. We first consider the case where $0<\phi_0<\pi/4$. In this case, as $\gamma$ increases, we expect that for a fixed $\phi_0$, the symmetry breaking will occur at $Q=\theta=0$. Therefore Eq.~(\ref{eq19}) can be rewritten as:
\begin{equation}
\label{eq20}
{cosh}^{2}(\gamma)-4cos(\phi_0)cosh(\gamma)+2{cos}^{2}(\phi_0)+1=0
\end{equation}
From here one can easily show that this critical $\gamma$ is given by:
\begin{equation}
\label{eq21}
\gamma={cosh}^{-1}\left (2cos(\phi_0)-\sqrt{cos(2\phi_0)}\right)
\end{equation}
This relation dictates the merging condition for the first two bands and is only valid for $0<\phi_0<\pi/4$, consistent with our previous observations. To find the corresponding relation for the band merging occurring at the edges, in Eq.~(\ref{eq19}) we set $Q=\pi/2$ , which in turn leads to a second order algebraic equation in $cos(\theta)$. Since we expect that the two eigenvalues will collapse into one (exceptional point), one may use this degeneracy condition in Eq.~(\ref{eq19}) at $Q=\pi/2$. After setting the discriminant of the quadratic equation to zero one finds that:
\begin{equation}
\label{eq22}
\gamma={cosh}^{-1}\left (\sqrt{2}\right )\approx 0.8814
\end{equation}
This last relation provides the $\mathcal{PT}$-threshold for band merging at the edges of the Brillouin zone and is independent of $\phi_0$. Interestingly this same value $\gamma={cosh}^{-1}(\sqrt{2})$ coincides with the critical $\mathcal{PT}$-threshold of the basic unit involved in this lattice, as found in section V.\\

\begin{figure}[t,b,h]
%\begin{center}
\includegraphics[width=3.5in]{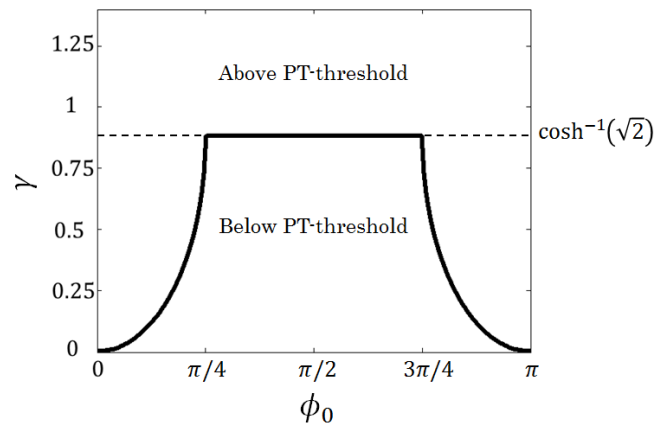}
\caption{(Color online) $\mathcal{PT}$-symmetry breaking threshold curve in a two dimensional parameter space of $\phi_0$ and $\gamma$. The region below the curve corresponds to the exact PT-phase while the region above the curve designates the domain where $\mathcal{PT}$ symmetry is broken.}
\label{fig13}
%\end{center}
\end{figure}

Figure~\ref{fig13} depicts the $\mathcal{PT}$-symmetry breaking threshold in the parameter space of $\phi_0$ and $\gamma$. The area below the curve corresponds to the case where the system operates in the exact $\mathcal{PT}$ phase where all the eigenvalues are real. On the curve symmetry breaking occurs and above this line the spectrum is in general complex. The top flat line of this curve corresponds to the critical value of $0.8814$ while the part between $0<\phi_0<\pi/4$ can be obtained from Eq.~(\ref{eq21}). The other segement symmetrically follows.

To dynamically explore the symmetry breaking threshold, the impulse response of our system is studied. Since the impulse is expected to excite the entire band of this mesh lattice, one should expect that an exponential growth in the total energy of the system should be observed once the $\mathcal{PT}$-symmetry is broken.

\begin{figure}[t,b,h]
%\begin{center}
\includegraphics[width=3.5in]{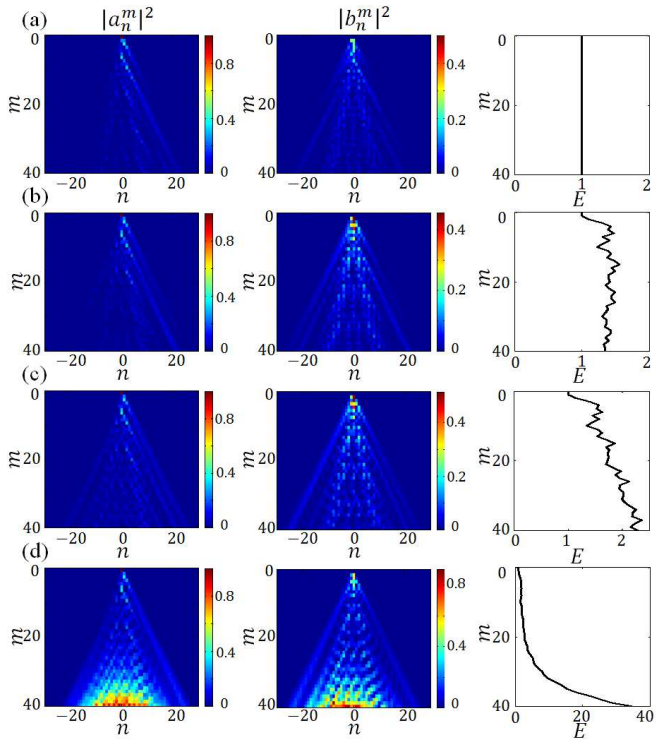}
\caption{(Color online) Impulse response of the $\mathcal{PT}$-symmetric lattice with a periodic phase potential of $\phi_0=0.2\pi$ while several different amounts of gain/loss are considered; (a) $\gamma=0$ (the passive lattice), (b) $\gamma=0.3$ (below threshold), (c) $\gamma=0.35$ (at threshold), (d) $\gamma=0.4$ (above threshold)}
\label{fig14}
%\end{center}
\end{figure}

Figure~\ref{fig14} shows the impulse response ($a_n^0=1$ , while all other elements are initially zero) of a $\mathcal{PT}$-symmetric mesh lattice for several different values of gain/loss $\gamma$ when $\phi_0=0.2\pi$. This range covers the passive scenario, or the case where the system operates below, at, and above the $\mathcal{PT}$-symmetry breaking threshold. The total energy in the system $E_m=\sum_{n}{\left|a_n^m\right |}^2+{\left|b_n^m\right |}^2$, is also plotted in each case at each discrete step of propagation, $m$ in Fig.~\ref{fig14}. While for the passive system ($\gamma=0$) the total energy remains constant during propagation, for a $\mathcal{PT}$-symmetric lattice used below its threshold the total energy tends to oscillate during propagation -but always remains below a certain bound. Note that such power oscillations were previously encountered in other $\mathcal{PT}$-symmetric periodic structures \cite{a16}. At exactly the $\mathcal{PT}$-threshold a linear growth in energy is observed-see Fig.~\ref{fig14}(c). Finally above threshold an exponential growth in energy is observed as expected from a system involving complex eigenvalues- Fig.~\ref{fig14}(d).\\

To further explore the behavior of this $\mathcal{PT}$-synthetic mesh lattice, we use at the input a Gaussian wavepacket, as in Eq.~(\ref{eq12}). Indeed by exciting this system with a wide input beam (that has a narrow spectrum) one can selectively excite different sections of the band structure. We now consider a $\mathcal{PT}$-symmetric mesh lattice with a periodic phase potential of amplitude $\phi_0=0.2\pi$ and a gain/loss factor of $\gamma=0.4$. The band structure corresponding to this structure is plotted in Fig.~\ref{fig12}.

\begin{figure}[t,b,h]
%\begin{center}
\includegraphics[width=3.5in]{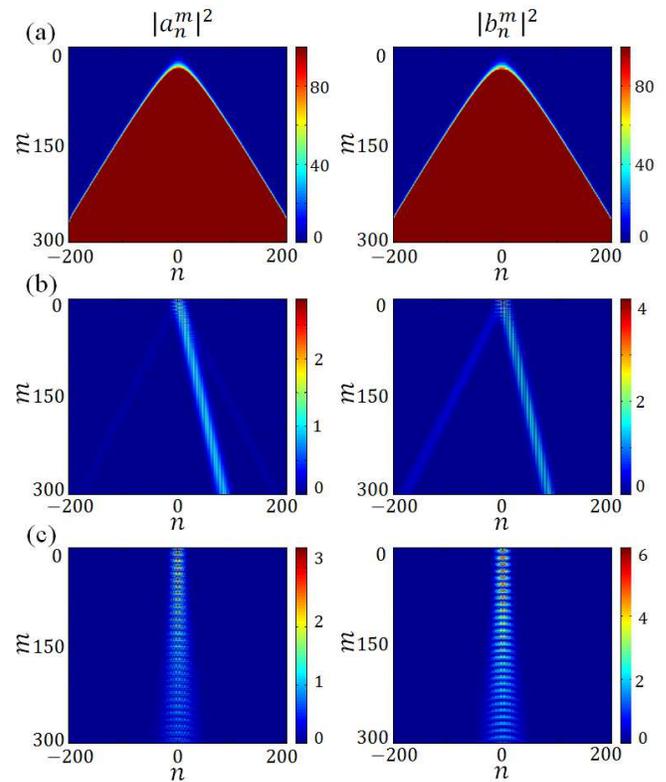}
\caption{(Color online) Gaussian beam propagation in a $\mathcal{PT}$-symmetric lattice operating in the broken $\mathcal{PT}$ phase regime. The lattice has a periodic phase potential of amplitude $\phi_0=0.2\pi$ and a gain/loss factor of $\gamma=0.4$. The Gaussian beam has a width of $2\Delta=30$ and is launched with three different values of initial phase tilt; (a) $Q_0=0$, (b) $Q_0=0.25\pi$, (c) $Q_0=0.5\pi$. In (a) the intensities are only shown up to a level of 100.}
\label{fig15}
%\end{center}
\end{figure}

Figure~\ref{fig15} depicts the propagation of a Gaussian wavepacket in this lattice, when launched with a Bloch momentum $Q_0$. Three different values for $Q_0$ have been selected for this example: $Q_0=0$, $0.25\pi$ and $0.5\pi$. According to Fig.~\ref{fig15} while for the first case an exponential energy growth is observed, for the other two cases energy remains essentially bounded. These results reveal that even above the $\mathcal{PT}$-symmetry breaking threshold, non-growing/decaying modes can be excited in such systems. This all depends on which section of the band structure is excited by the initial conditions.\\

Compared to a passive mesh lattice, the band structure of its $\mathcal{PT}$-symmetric counterpart reveals another interesting property. As previously discussed, the maximum beam transport angle (${\Omega}_{max}$) in an empty lattice is $1/\sqrt{2}$, and even in the presence of a periodic phase potential this angle is always less than this maximum transverse velocity. However according to the Fig.~\ref{fig12}, when approaching the exceptional points from the real section (blue part) of the band, its slope tends to considerably increase and eventually approaches infinity around the exceptional points.

\begin{figure}[t,b,h]
%\begin{center}
\includegraphics[width=3.5in]{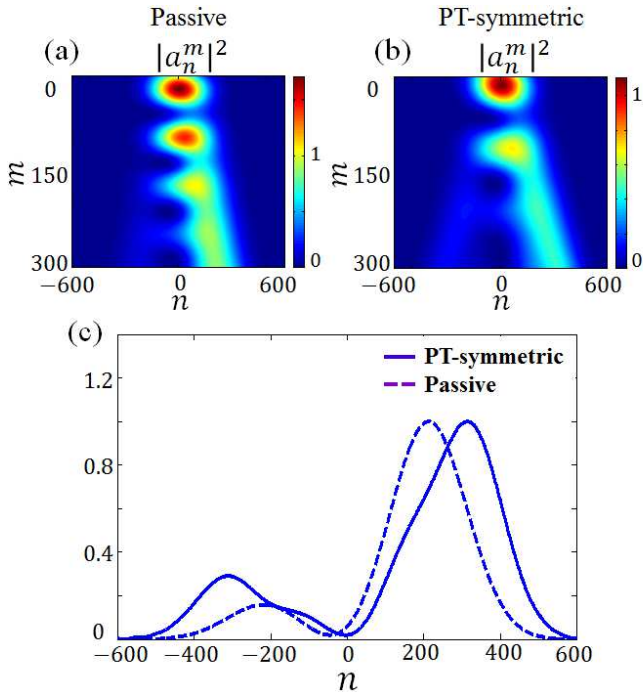}
\caption{(Color online) A broad Gaussian beam propagating in a passive and a $\mathcal{PT}$-symmetric lattice; (a) evolution of the Gaussian beam in a passive empty lattice, (b) in a $\mathcal{PT}$-symmetric lattice, (c) normalized intensity profiles of the beam at the last propagation step ($m=300$) in both lattices. The parameters of the $\mathcal{PT}$ lattice are $\gamma=0.039$ and $\phi_0=0$. The Gaussian beam has a beam width of $2\Delta=400$ and an initial phase front tilt of $Q_0=0.9817\pi$.}
\label{fig16}
%\end{center}
\end{figure}

Figure~\ref{fig16} compares the propagation of a Gaussian beam in a passive and a $\mathcal{PT}$-symmetric mesh lattice operating above threshold. Both lattices are excited with the same Gaussian beam having a Bloch momentum $Q_0$, which is chosen to be close to the exceptional point of the $\mathcal{PT}$-symmetric lattice. Close to this exceptional point, the slope of the band structure tends to infinity and therefore, the associated group velocity can become almost arbitrarily high for any narrow-bandwidth wavepacket. While the maximum beam angle in passive empty lattice is $\sim 0.7$ (which is close to the maximum) for the $\mathcal{PT}$-symmetric lattice this angle is approximately $1.04$ which is certainly above the maximum limit of the passive lattice. This effect has in fact a counterpart in continuous media. As previously shown, in the presence of a gain medium \cite{a48,a49} and in $\mathcal{PT}$-symmetric and gain/loss gratings and lattices \cite{a36,a491,a50} used close to the exceptional points, the group velocity of light can be superluminal. It should be noted however that none of these effects violates causality since non causal waveforms are used for excitation. Indeed, this superluminal propagation of the intensity peak is enabled by a gain-assisted growth of the distribution's tails.\\

Finally we investigate the concept of unidirectional invisibility in a $\mathcal{PT}$-symmetric mesh lattice. As recently predicted \cite{a33,a34} $\mathcal{PT}$-symmetric periodic structures like gratings can exhibit surprising behavior like unidirectional invisibility and intriguing reflection characteristics. More specifically, light propagating in such a system can experience reduced or enhanced reflections depending on the direction of propagation. Even more remarkable is what happens right at $\mathcal{PT}$ threshold: in this case light waves entering this arrangement from one side do not experience any reflection and can fully traverse the grating with unity transmission. Given that this occurs without acquiring any phase imprint from this $\mathcal{PT}$ system, the periodic structure is essentially invisible. Similar effects also occur in $\mathcal{PT}$-symmetric mesh lattices. This can be demonstrated in a system where one lattice is embedded in another lattice. This is achieved using for example the following phase modulation potential:
\begin{equation}
\label{eq23}
\phi_{n}=
\begin{cases}
~0,~~~~~~|n|>n_0\\
+\phi_{0},~~~|n|<n_0,~n:~ even\\
-\phi_{0},~~~|n|<n_0,~n:~~ odd
\end{cases}
\end{equation}
where $2n_0$ is the width of this $\mathcal{PT}$ grating lattice. In this same way the anti-symmetric gain/loss profile is also imposed only in the grating region $|n|<n_0$.\\

\begin{figure}[t,b,h]
%\begin{center}
\includegraphics[width=3.5in]{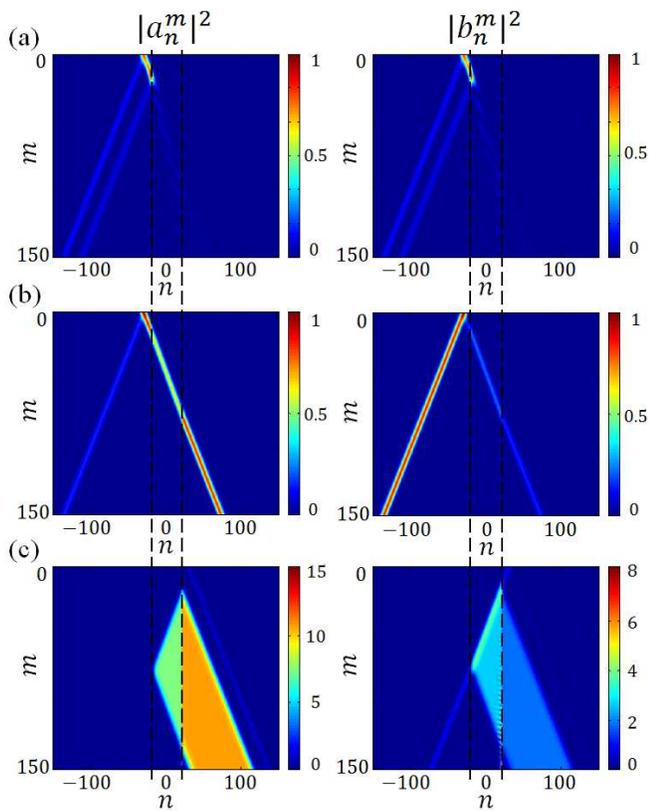}
\caption{(Color online) Unidirectional scattering of a Gaussian beam from a $\mathcal{PT}$-symmetric grating. A $\mathcal{PT}$-symmetric grating having $40$ layers with a phase potential of $\phi_0=0.2\pi$ and a gain/loss of $\gamma=0.3507$ (corresponding to $\mathcal{PT}$ threshold) is established in the middle of an empty lattice. The dashed line shows the grating region. (a) Scattering from the passive grating, (b) left side scattering from this $\mathcal{PT}$-symmetric grating, (c) right side scattering from this same grating.}
\label{fig17}
%\end{center}
\end{figure}

To investigate this latter process, a $\mathcal{PT}$-symmetric lattice having $2n_0=40$ layers is embedded inside an empty lattice. The empty lattice is excited with a Gaussian beam (as in Eq.~(\ref{eq12})) with an initial phase tilt of $Q_0=\pi$. Scattering of the beam from the left and right side of this $\mathcal{PT}$-symmetry grating is depicted in Fig.~\ref{fig17}.

Figure~\ref{fig17} shows the scattering of Gaussian wavepacket from this grating when $2n_0=40$, $\phi_0=0.2\pi$, and $\gamma=0.3507$. The extent of this grating is shown by the dashed lines. Figure~\ref{fig17}(a) shows how the Gaussian beam is scattered or transmitted by this grating when $\gamma=0$. The effects of reflection and reduced transmission are evident in this figure. In Fig.~\ref{fig17}(b) we show these same dynamics for $\gamma=0.3507$ and when the Gaussian excites the grating from the left. Both the $a_n^m$, $b_n^m$ channels are excited during this process. In this case no reflections occur when the grating is used close to the exceptional point. Essentially in this regime the grating leaves no mark on the beam itself and hence is practically invisible. Note that the associated splinter beams in this figure do not represent reflections-they simply emerge from the two different bands associated with the empty lattice. On the other hand, Fig.~\ref{fig17}(c) shows what will happen when the Gaussian beam excites the right side of this same $\mathcal{PT}$-symmetric lattice grating. In this latter case pronounced reflections occur (even exceeding unity) and the grating ceases to be invisible to light.\\

\section{CONCLUSIONS}

In this work we have studied the properties of a new class of periodic structures both in the passive as well in the $\mathcal{PT}$-symmetric regime. These optical mesh lattices are in essence waveguide arrays that are discretely and periodically coupled to each other along the propagation direction. In addition phase elements can also be used in appropriate positions to control the phases while amplifiers/attenuators can be employed to realize the antisymmetric imaginary part of the $\mathcal{PT}$ potential. What makes these optical lattices different from previously known waveguide array versions is the presence of discreetness in both the transverse and longitudinal directions. The band structure of these systems has been systematically analyzed and an analytic expression was obtained for their dispersion relation. We have shown that the band structure is periodic in both the propagation constant and transverse Bloch momentum while the Brillouin zone of these lattices displays in general four bands. In addition we found that the shape of the bands and band gaps can be effectively controlled using phase elements. Interestingly, through a proper phase modulation the band structure can be arbitrary shifted from the center. It should be noted that shifting the bands in standard optical waveguide array systems is not straightforward and typically requires the presence of external magnetic field effects. The impulse light dynamics as well as wavepacket excitations were numerically explored and related to the properties of the band structure. The elementary $\mathcal{PT}$-building block involved in these mesh arrangements was examined and its symmetry breaking threshold was determined. As it was discussed, what could greatly facilitate the physical realization of such a large scale $\mathcal{PT}$-symmetric mesh lattice is the fact that couplings and amplification/attenuation can be independently controlled within the basic building block of this lattice. Band merging effects in these lattices were investigated and the conditions for spontaneous $\mathcal{PT}$-symmetry breaking were explicitly obtained in terms of relevant parameters. The response of these systems under impulse and broad beam excitation was investigated in terms of their respective band structure. As it was shown, light dynamics in the $\mathcal{PT}$-symmetric lattice exhibits certain peculiarities that are otherwise impossible in its passive counterpart. These include for example power oscillations and transitions from neutral to exponentially growing regimes. The possibility of superluminal transport along with unidirectional invisibility was also considered.\\

\begin{acknowledgments}

This work has been partially supported by NSF Grant No. ECCS-1128520 and AFOSR Grant No. FA95501210148. It was also funded by DFG Forschergruppe 760, the Cluster of Excellence Engineering of Advanced Materials and School of Advanced Optical Technologies (SAOT).

\end{acknowledgments}
\bibliography{GF}
\end{document}